\documentclass[twocolumn,showpacs,preprintnumbers,amsmath,amssymb]{revtex4}

\usepackage{graphicx}% Include figure files
\usepackage{dcolumn}% Align table columns on decimal point
\usepackage{bm}% bold math
\begin{document}
%--------------------------------------------------------------------
\title{Rheology of High-Capillary Number Flow in Porous Media}
%--------------------------------------------------------------------
\author{Santanu Sinha\email{santanu@csrc.ac.cn}}
\affiliation{Beijing Computational Science Research Center, 10 East
Xibeiwang Road, Haidian District, Beijing 100193, China.}
\author{Magnus Aa.\ Gjennestad,\email{magnus.a.gjennestad@ntnu.no}}
\affiliation{PoreLab, Department of Physics, Norwegian University of
Science and Technology, NO--7491 Trondheim, Norway.}
\author{Morten Vassvik,\email{morten.vassvik@ntnu,no}} 
\affiliation{PoreLab, Department of Physics, Norwegian University of
Science and Technology, NO--7491 Trondheim, Norway.}
\author{Mathias Winkler\email{mathias.winkler@ntnu.no}}
\affiliation{PoreLab, Department of Physics, Norwegian University of
Science and Technology, NO--7491 Trondheim, Norway.}
\author{Alex Hansen\email{alex.hansen@ntnu.no}}
\affiliation{PoreLab, Department of Physics, Norwegian University of
Science and Technology, NO--7491 Trondheim, Norway.}
\author{Eirik G.\ Flekk{\o}y\email{e.g.flekkoy@fys.uio.no}}
\affiliation{PoreLab, Department of Physics, University of Oslo,
  P.O.\ Box 1048 Blindern, NO-0316 Oslo, Norway}
%--------------------------------------------------------------------
\date{\today {}}
%--------------------------------------------------------------------
\begin{abstract}
Immiscible fluids flowing at high capillary numbers in porous media
may be characterized by an effective viscosity.  We demonstrate that
the effective viscosity is well described by the Lichtenecker-Rother
equation. The exponent $\alpha$ in this equation takes either the 
value 1 or 0.6 in two- and 0.5 in three-dimensional systems depending on the pore
geometry. Our arguments are based on analytical and numerical
methods.
\end{abstract}
%--------------------------------------------------------------------
\maketitle
%--------------------------------------------------------------------

The hydrodynamics of real things very often happens at small scale,
i.e.~in a porous medium \cite{b72}. This is the case in a wide variety
of biological, geological and technological systems where there are
normally several immiscible fluids present.  The challenge of
describing such systems in a unified way, however, is largely
unsolved. An important reason for this is the lack of a length scale
above which the system may be averaged as if it were homogeneous. Such
a length scale gives rise to the so-called representative elementary
volume (REV) which is the conceptual basis for conventional theories
that seek to up-scale the description of flow in porous
media. However, since the fluid structures in question are often
fractal, the REV average of intensive quantities, such as saturations,
will depend on the size of the REV.

An important and rather general exception where this is not a problem,
is the case of steady states \cite{tkrlmtf09,aetfhm14}. Steady states
are characterized by potentially strong fluctuations at the pore
scale, but with steady averages at the REV scale. As such they differ
fundamentally from stationary states that are static at the pore scale
as well. Steady states have much in common with ensembles in
equilibrium statistical mechanics. These states are implicitly assumed
in conventional descriptions of porous media flows that take the
existence of an REV for granted.

When the flows in question contain immiscible phases that are strongly
forced in the sense that viscous forces dominate capillary forces, the
description of the steady state simplifies to the description of a
single fluid. This is the subject of the present letter, and we show
how the emergent description is manifestly incompatible with the
conventional theories that have been in use for more than 80 years,
most notably perhaps by the petroleum industry.

The first and still leading theory describing immiscible two-phase
flow in porous media is that of Wyckoff and Botset \cite{wb36} who 
based their theory of relative permeability on the idea that when
the porous medium is seen from the viewpoint of one of the fluids, the
pore volume accessible to this fluid would be the pore volume of the
porous medium minus the pore volume occupied by the other fluid.  This
reduces the effective permeability seen by either fluid and the
relative reduction factor is the relative permeability.  In order to
account for the surface tension between the immiscible fluids in the
pores, the concept of capillary pressure was introduced by Leverett \cite{l40}.
The central equations in relative permeability theory are the
generalized Darcy equations
\begin{equation}  
\label{eq050}
{\vec v}_j=-\frac{K}{\mu_j}\ k_{r,j}(S_j)\  {\vec\nabla}P_j\;,
\end{equation}
where the subscript $j$ either refers to the wetting fluid ($j=w$) or
the non-wetting fluid ($j=n$).  ${\vec v}_w$ and ${\vec v}_n$ are
superficial velocities of the wetting and non-wetting fluids,
respectively. They are defined as the volumetric flow rates of each
fluid entering a REV divided by the area of entry.
$K$ is the permeability of the porous medium, $\mu_w$ and $\mu_n$ are
the wetting and non-wetting viscosities. $k_{r,w}(S_w)$ and
$k_{r,n}(S_w)$ are the relative permeabilities of the wetting and
non-wetting fluids and they are both functions of the wetting
saturation $S_w$ only.  The corresponding non-wetting saturation is
$S_n$ and we have that $S_w+S_n=1$.  The wetting and non-wetting
pressure fields $P_w$ and $P_n$ are related through the capillary
pressure function $P_c(S_w)=P_n-P_w$.
We may define a total superficial velocity ${\vec v}$  
\begin{equation}
\label{eq90}
{\vec v}={\vec v}_w+{\vec v}_n\;.
\end{equation}
This is the volumetric flow rate of all fluids entering the REV divided by
the area of entry. 

Let us now suppose that the flow rates are so large that the capillary
pressure may be ignored.  Hence, we have $P_n=P_w=P$ and we may
combine the relative permeability equations (\ref{eq050}) with
equation (\ref{eq90}) to find
\begin{equation}
\label{eq095}
{\vec v}=-{K}\left[\frac{k_{r,w}(S_w)}{\mu_w}
+\frac{k_{r,n}(S_n)}{\mu_n}\right]{\vec\nabla}P
=-\frac{K}{\mu_{\rm eff}(S_w)}{\vec\nabla}P\;,
\end{equation}
where we have defined an 
{\it effective viscosity\/} $\mu_{\rm eff}$
\begin{equation}
\label{eq096}
\frac{1}{\mu_{\rm eff}(S_w)}=
\frac{k_{r,w}(S_w)}{\mu_w}+\frac{k_{r,n}(S_n)}{\mu_n}\;.
\end{equation}

There have been many suggestions as to what functional form the
relative permeabilities $k_{r,w}(S_w)$ and $k_{r,n}(S_w)$ take.  The
most common choice is to use the Brooks--Corey relative permeabilities assuming
$k_{r,w}(S_w)=k^0_{r,w} S_w^{n_w}$ and $k_{r,n}(S_w)=S_n^{n_n}$ where $0\le
k^0_{r,w}\le 1$  and the two Corey exponents $n_w$ and $n_n$ are typically in the range
2 to 6 \cite{bc64,l89}.

Equation (\ref{eq096}) is problematic. When $\mu_w=\mu_n$, a dependency of $\mu_{\rm eff}$ 
on the saturation is predicted when $n_w$ and/or $n_n$ are larger than 1 when using
the Brook--Corey relative permeabilities. Other functional forms for the relative
permeabilities give similar dependencies.  Clearly, such behavior is not physical.

McAdams et al.\ \cite{mwh42} proposed an effective viscosity for
two-phase flow by assuming a saturation-weighted {\it harmonic\/}
average
\begin{equation}
\label{eq100}
\frac{1}{\mu_{\rm eff}}=\frac{S_w}{\mu_w}+\frac{S_n}{\mu_n}\;.
\end{equation}
On the other hand, Cicchitti et al.\ \cite{clssz60} proposed an
effective viscosity based on the saturation-weighted {\it
  arithmetic\/} average
\begin{equation}
\label{eq200}
\mu_{\rm eff}=\mu_w S_w+\mu_n S_n\;.
\end{equation}
Both of these expressions become saturation-independent when
$\mu_w=\mu_n$ as they should. There are several other proposals for
the functional form of the effective viscosity $\mu_{\rm eff}$ in the
literature \cite{am08}.

A one-dimensional porous medium, i.e.\ a capillary tube where the two
fluids move as bubbles in series \cite{shbk13} constitutes a series
coupling and the arithmetic average (\ref{eq200}) is the
appropriate one.  On the other hand, if the capillary tubes forms a
parallel bundle each filled with either only the wetting or the
non-wetting fluid, we have a parallel coupled system and equation
(\ref{eq100}) is appropriate.  Suppose now that each capillary $i$
in the bundle is filled with a bubble train with a corresponding
wetting saturation $S_{w,i}$.  The probability distribution for
finding a capillary having this saturation, $S_{w,i}$, is
$p(S_{w,i})$.  We have that
\begin{equation}
\label{eq300}
S_w=\int_0^1\ dS\ p(S)\ S\;.
\end{equation}
The effective viscosity for the capillary bundle is then given by
\begin{equation}
\label{eq400}
\frac{1}{\mu_{\rm eff}}
=\int_0^1 \frac{p(S)\ dS}{\mu_w S+\mu_n(1-S)}\;.
\end{equation}
As a model for the distribution $p(S_{w,i})$, we may take a gaussian
with a narrow width $\sigma$ centered around $S_w$:
$p(S_{w,i})=\exp[-(S_{w,i}-S_w)^2/2\sigma^2]/\sqrt{2\pi \sigma^2}$.
Using a saddle point approximation, we find to order $\sigma^2$ that
\begin{equation}
\label{eq500}
\mu_{\rm eff}= \mu_w S_w+\mu_n S_n 
-\frac{(\mu_n-\mu_w)^2}{\mu_w S_w+\mu_n S_n}\ \sigma^2\;.
\end{equation}  

We now consider a wide distribution of saturations in each capillary:
$p(S_{w,i})$ is uniform rather than gaussian.  We set the average
wetting saturation to the value $S_w=1/2$, finding
\begin{equation}
\label{eq525}
\mu_{\rm eff}=\left|\frac{\mu_w-\mu_n}
   {\ln\left(\frac{\mu_w}{\mu_n}\right)}\right|\;.
\end{equation}
The functional form of the latter equation is very different from the
gaussian, equation (\ref{eq500}).

The extreme case when the capillaries are either filled completely by
the wetting or the non-wetting fluid corresponds to
$p(S_{w,i})=S_w\delta(S_{w,i}-1)+S_n\delta(S_{w,i})$ giving, as
already pointed out, an effective viscosity according to equation
(\ref{eq100}).  We may, however, study this either-or situation in a
more complex network, namely a square lattice.  We assume that the
wetting saturation is set to $S_w=1/2$, which defines the percolation
threshold for bond percolation and that the links are randomly filled
with either fluid.  By analogy with the percolation problem we may use
Straley's exact result \cite{s77} leading to an effective viscosity
given by
\begin{equation}
\label{eq550}
\mu_{\rm eff}=\sqrt{\mu_w\mu_n}\;.
\end{equation}   

We may calculate the effective viscosity of a regular lattice by using
Kirkpatrick's mean field theory \cite{k73}, first introduced as a tool
to calculate the conductivity of a percolating set of conductors with
random values. In that case conductances are fixed in time and space,
giving rise to a problem different from ours where the fluids are free
to distribute themselves in a non-trivial way and create flow paths
where conductances vary dynamically.

The mobility between nodes $i$ and $j$ is $K_{ij}/\mu_{ij} $ where
$K_{ij}$ is the permeability and $\mu_{ij} = \mu_w S_{w,ij} + \mu_n
S_{n,ij} $ is the effective viscosity of the link and the saturations
take on their local values. The form of $\mu_{ij}$ reflects the fact
that the fluids contained in the link are connected in series.

Kirkpatrick's theory is based on the notion that the network of
mobilities $K_{ij}/\mu_{ij}$ may be replaced by a network of links
with a single mobility $K/\mu_{\rm eff}$ but with the same total
network mobility. The value of $K/\mu_{\rm eff}$ is given by the
formula \cite{k73}
\begin{equation} 
\label{eq800}
\left\langle 
\frac{\frac{K}{\mu_{\rm eff}} - 
\frac{K_{ij}}{\mu_{ij}}}{\frac{K_{ij}}{\mu_{ij}} + 
\left[\left(\frac{z}{2}\right)%
 - 1\right]\frac{K }{\mu_{\rm eff}}} 
\right\rangle=0\;,  
\end{equation}
where $z$ is the coordination number of the lattice, and the ensemble
average $\langle ... \rangle = \int_0^\infty dK_{ij} P(K_{ij})%
\int_0^1 dS_{w,ij} p(S_{w,ij}) ...$, $P(K_{ij})$ being the
permeability distribution and $p(S_{w,ij})$ is the wetting saturation
distribution fulfilling (\ref{eq300}).  We assume a square lattice
so that $z=4$.

By assuming that the saturation distribution is a narrowly peaked
gaussian with width $\sigma$, we may again use the saddle point 
approximation giving
\begin{equation}
\label{eq850}
\mu_{\rm eff}= \mu_w S_w+\mu_n S_n +{\cal O}\left(|\mu_n-\mu_w|\sigma^2\right)\;.
\end{equation}
This result is similar to that found for the parallel capillary
bundle, see equation (\ref{eq500}).

From these model systems giving rise to equations (\ref{eq500}), 
(\ref{eq525}), (\ref{eq550}) and 
(\ref{eq850}) we see that it is far from obvious what the effective
viscosity $\mu_{\rm eff}$ should be.  Does it depend on the details of
the porous medium or can one find a general form?  We may generalize
equations (\ref{eq100}) and (\ref{eq200}) by writing them in the form
\begin{equation}
\label{eq600}
\mu_{\rm eff}^\alpha=\mu_w^\alpha S_w+\mu_n^\alpha S_n\;,
\end{equation}
where $\alpha=-1$ for parallel coupling --- equation (\ref{eq100}) ---
and $\alpha=+1$ for series coupling --- equation (\ref{eq200}).  The
effective viscosity in (\ref{eq550}) corresponds to $\alpha=0$, 
whereas equations (\ref{eq500}) and (\ref{eq850}) suggest $\alpha=1$. Only
the effective viscosity in equation (\ref{eq525}) does not fit this form.
Equation (\ref{eq600}) has been used for estimating the effective
electrical permittivity of heterogeneous conductors and in connection
with permeability homogeneization in porous media. It is known as the 
Lichtenecker--Rother equation \cite{lr31,gp94,ts05,bc10}.

We now proceed with numerical methods to test whether the proposed
form (\ref{eq600}) adequately describes the effective
viscosity.  We use two approaches;
dynamic pore-network modeling and Lattice Boltzmann modeling. In
the dynamic pore-network modeling \cite{amhb98}, the porous medium 
is represented by a network of links, connected at nodes, 
which transport two immiscible fluids separated by interfaces. We implement 
periodic boundary conditions, which keeps the saturation constant with time. 
The flow rate of the fluids inside a link between two neighboring nodes 
$i$ and $j$ obeys the constitutive equation
\begin{equation}
\label{wbeq}
\displaystyle
q_{ij}= -\frac{g_{ij}}{l_{ij}}\left[p_j-p_i\right]\;,
\end{equation}
where $p_i$ and $p_j$ are the local pressure drops at the nodes. Here
$l_{ij}$ and $g_{ij}$ are respectively the length and the mobility of
the link. There is no contribution to the pressure from interfaces in 
the link as the surface tension has been set to zero. The mobility $g_{ij}$ 
is inversely proportional to the link viscosity given by 
$\mu_{ij}=\mu_w S_{w,ij}+\mu_n S_{n,ij}$.  Simulations are performed with 
a constant global pressure drop $\Delta P$ across the network. We determine 
the local pressures $p_i$ by solving the Kirchhoff equations using the 
conjugate gradient algorithm. Local flow rates $q_{ij}$ through each link are
then calculated using equation (\ref{wbeq}) and the interfaces are 
advanced with appropriate time steps.

A crucial point is the distribution of the two fluids after they
mix at the nodes and enter the next links. Whether the system allows
high or low fragmentation of the fluids, will depend on the geometry
of the porous media and the nature of the pore space  \cite{obl07,lzy16}. 
This will have impact on the size of the bubbles and the number of interfaces 
inside a link. Note that small bubbles of either fluid may not necessarily 
imply a large number of interfaces or vice-versa. We therefore implemented 
two different algorithms for the interface dynamics. In the {\it bubble-controlled\/}
algorithm, we
limit the minimum size of a bubble before entering in a link and in
the {\it interface-controlled\/} algorithm we limit the maximum number of
interfaces that can exist in a link. We then considered two different
possibilities for each algorithm, for the bubble-controlled case, (A)
bubbles with lengths at least equal to the respective pore radii, i.e
$b_{\rm min}\ge r_{ij}$ and (B) bubble sizes can be much smaller,
$b_{\rm min}\ge 0.02r_{ij}$. For the interface-controlled algorithm,
we studied two cases, (C) one with maximum 2 and (D) another with
maximum $4$ interfaces per link. For a detailed description of the two
algorithms for the interface dynamics, we refer to the supplementary
material.

We consider both two-dimensional (2D) and three-dimensional (3D)
networks. For 2D, we used square lattices of $64\times 64$ links
and honeycomb lattices of size $64\times 40$ links where all the links have
the same length $l$, and their radii $r_{ij}$ are uniformly
distributed in the interval $0.1\; l<r<0.4\; l$. The square network
was oriented at $45^\circ$ with respect to the overall flow
direction. For A and B, results of honeycomb lattices will be
presented, and for C and D, we will present the results of square
lattices. For 3D, two reconstructed pore networks extracted from the
real samples were used; one is a Berea sandstone containing $1163$
nodes and $2274$ links, and the another is a sand pack with $767$
nodes and $1054$ links \cite{sbdkpbtsch17}. The links for the 2D networks
have circular cross section which corresponds $g_{ij}=\pi
r_{ij}^4/(8\mu_{ij})$. For the 3D networks, the
links have triangular cross section and the mobility terms are
calculated considering their shape factors
\cite{sbdkpbtsch17,langglois64,jia08}.

\begin{figure}
\centerline{
  \includegraphics[width=0.35\textwidth,clip]{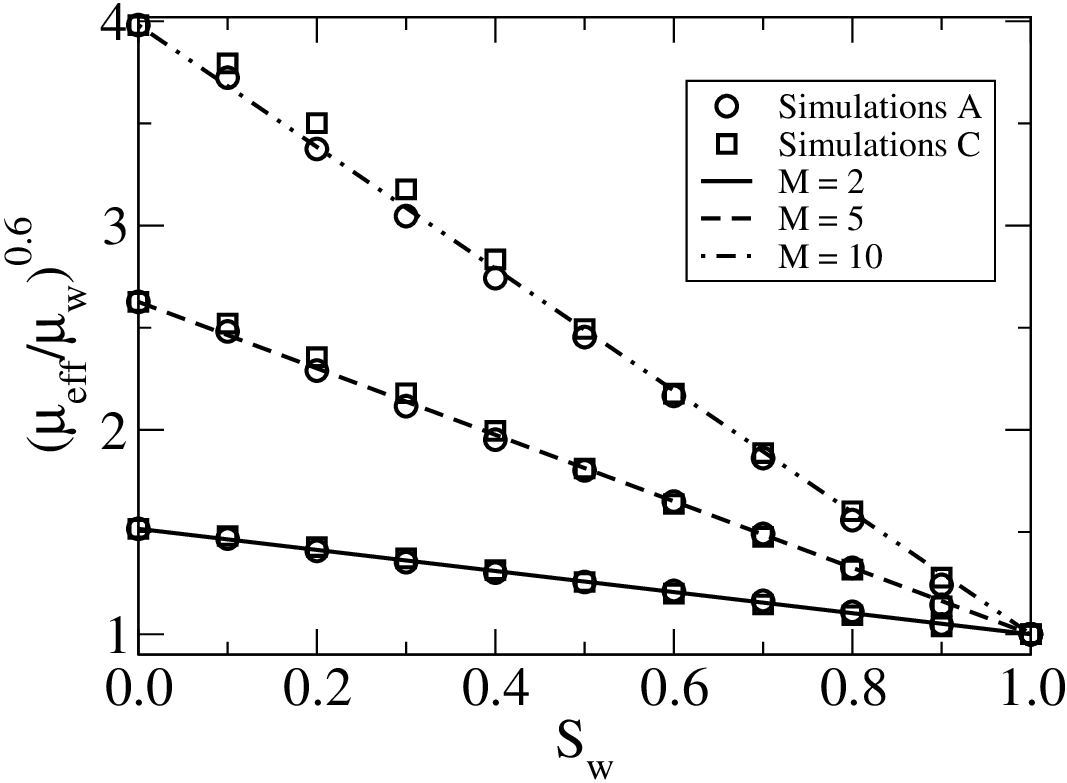}}
\centerline{
  \includegraphics[width=0.35\textwidth,clip]{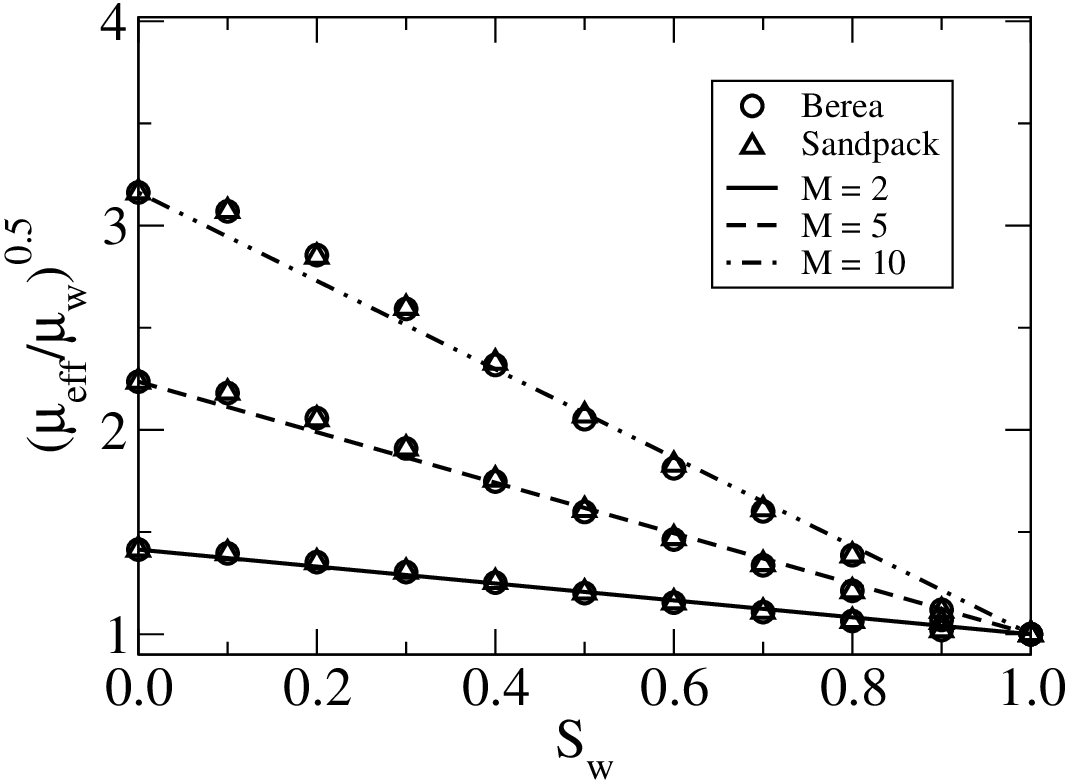}}
\caption{\label{fig1} Values of $(\mu_\text{eff}/\mu_\text{w})^\alpha$
  obtained from the network simulations for the cases A and C, the
  large bubbles and few interfaces respectively, plotted with
  different symbols as a function of the wetting saturations
  ($S_w$). Results are compared with $S_w+M^\alpha S_n$, equation
  (\ref{eq600}), plotted with straight lines, and found consistent
  with $\alpha = 0.6$ for two-dimensions and $\alpha = 0.5$ for three
  dimensions.}
\end{figure}

\begin{figure}
\centerline{
  \includegraphics[width=0.35\textwidth,clip]{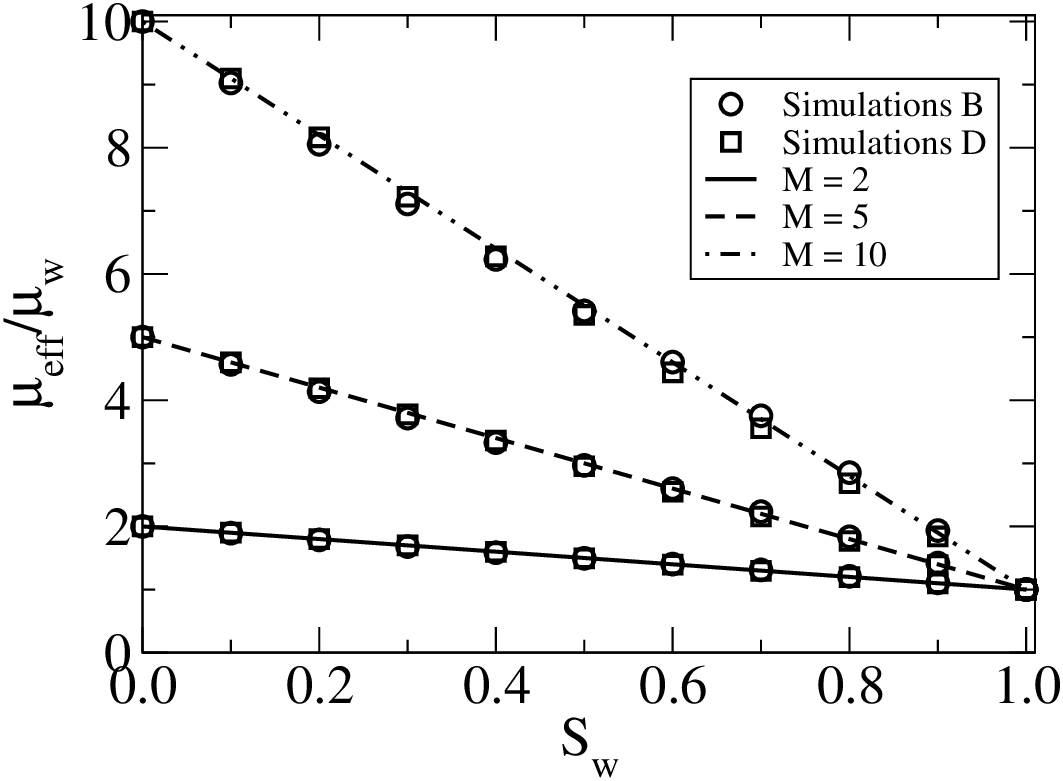}}
\centerline{
  \includegraphics[width=0.35\textwidth,clip]{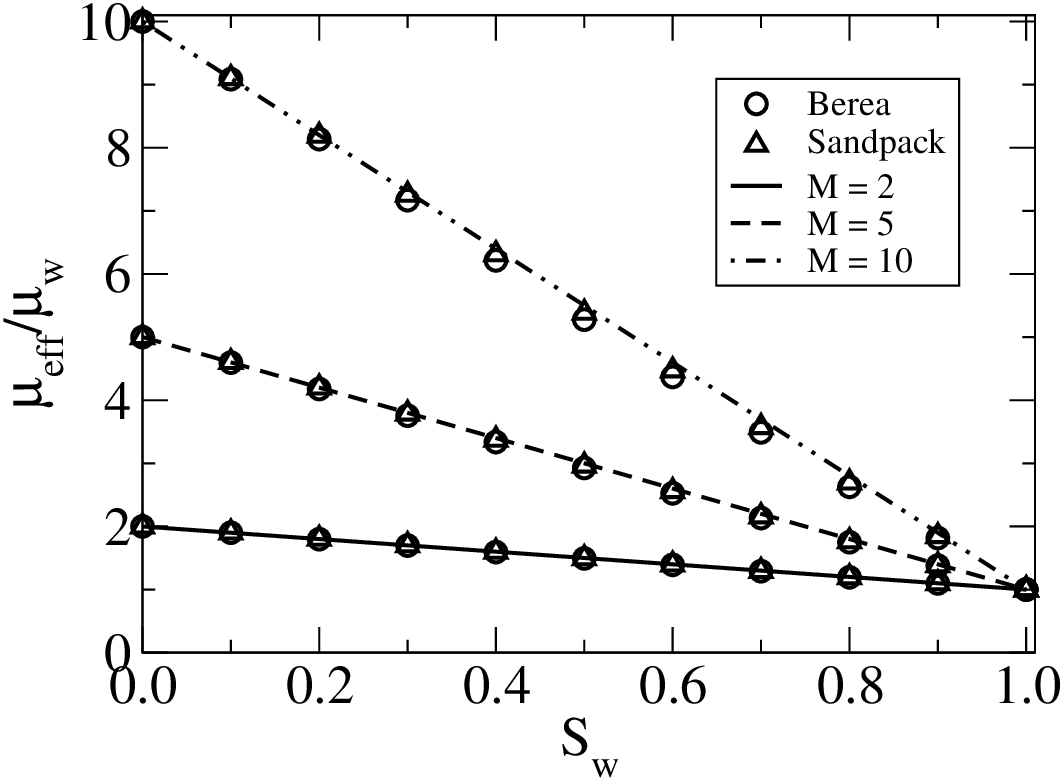}}
\caption{\label{fig2} Plot of $(\mu_\text{eff}/\mu_\text{w})^\alpha$
  obtained from the network simulations (symbols) for the cases B and
  D, small bubbles and many interfaces respectively, as a function
  of $S_w$ and compared with $S_w+M^\alpha S_n$ (straight lines). Here
  the numerical results are consistent with $\alpha = 1$ for both 2D
  and 3D.}
\end{figure}

Effective viscosities $\mu_{\rm eff}$ measured in
the steady state for models A and C (large bubbles or few interfaces) are
plotted in figure \ref{fig1}. Results are then compared with
$(\mu_{\rm eff}/\mu_w)^\alpha=S_w+M^\alpha S_n$, see equation (\ref{eq600}),
where $M=\mu_n/\mu_w$, and found consistent with $\alpha=0.6$ for 2D
and with $\alpha=0.5$ for 3D. In figure \ref{fig2}, we show 
$\mu_{\rm eff}/\mu_w$ for models B and D (small bubbles or many interfaces) and the
results show $\alpha=1$ for both 2D and 3D.  We show typical configurations
for models C and D in the supplementary material.

%-----------------------------------------------------------------
%-----------------------------------------------------------------
%-----------------------------------------------------------------
%-----------------------------------------------------------------

In summary we find for network models B and D, where there are
the minimum bubble size is 0.02 times the link radius (B), or up to 
four interfaces (D) a result which is consistent with the Kirkpatrick
mean field theory, $\alpha=1$. For the models with minimum bubble 
sizes being larger than the link radii (A) or with a maximum number
of interfaces equal to 2 (C), we find a much smaller $\alpha = 0.6$
in 2D and 0.5 in 3D (model C). Using the bubble-controlled model, we 
have varied the minimum bubble size over the range $0.02r_{ij}$
to $0.5r_{ij}$ finding $\alpha$ decreasing gradually from 1 to 0.6. 
Taking into account that $r_{ij}\le 0.4\; l$, where $l$ is the link length, 
this shift of $\alpha$ from 1 to 0.6 occurs over the the narrow 
range from $0.008\; l$ to $0.2\; l$, indicating that we are dealing with 
a crossover.     

\begin{figure}
\centerline{
  \includegraphics[width=0.35\textwidth,clip]{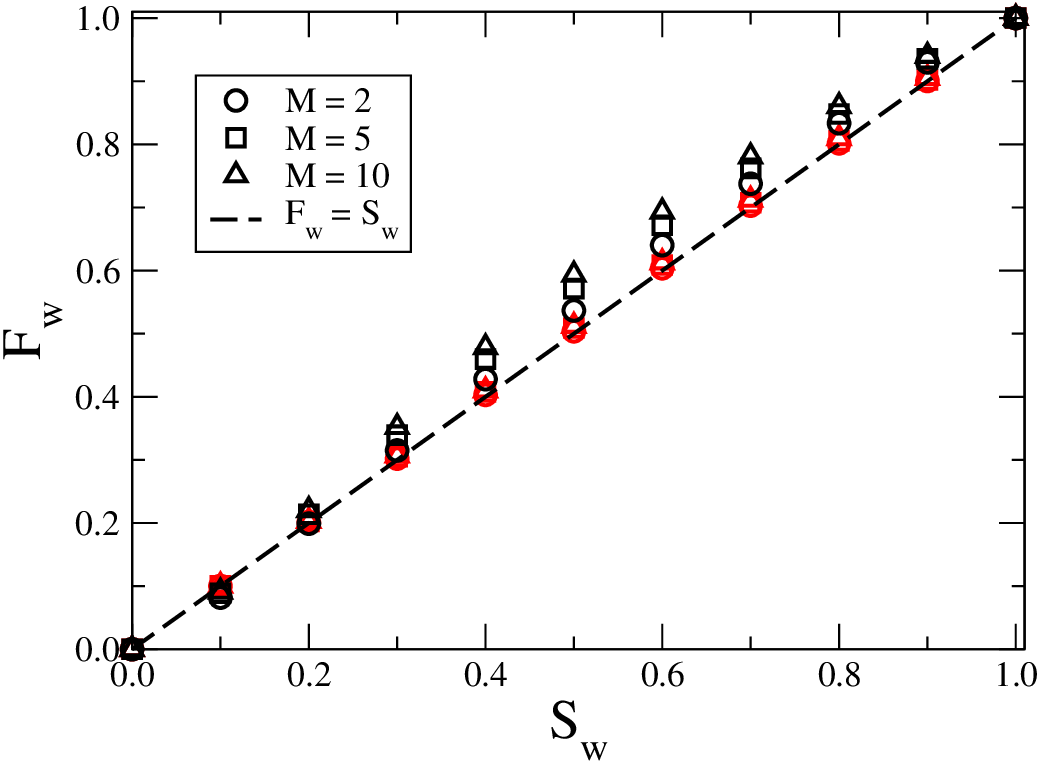}}
\centerline{
  \includegraphics[width=0.35\textwidth,clip]{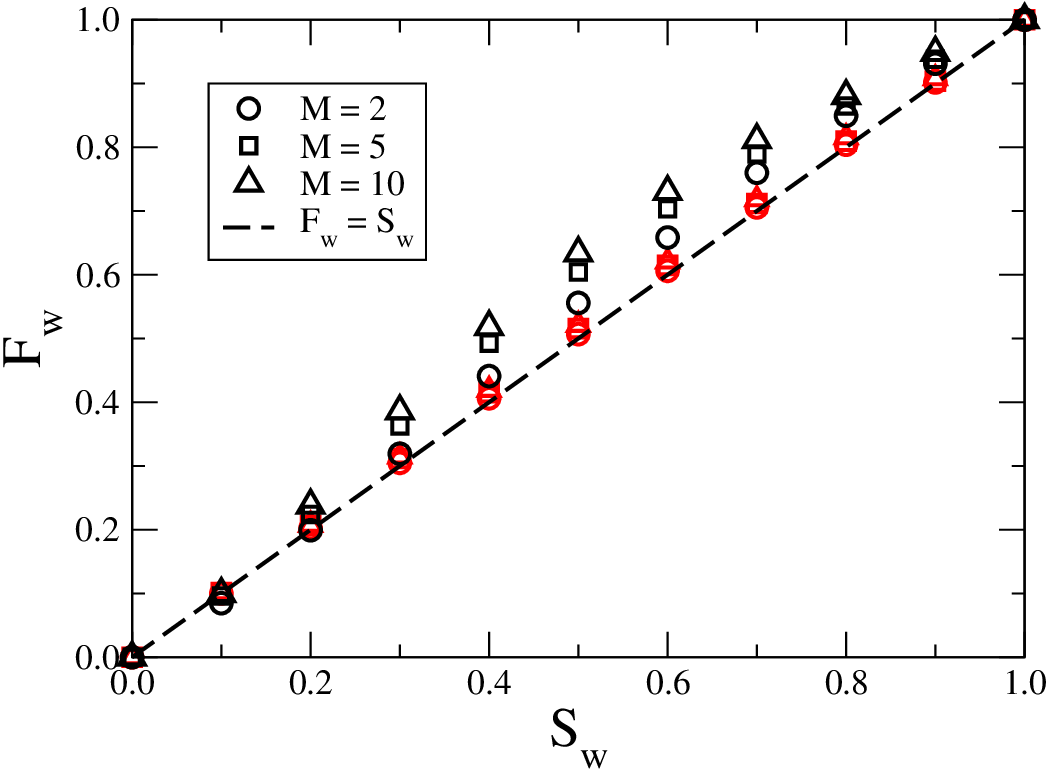}}
\caption{\label{fig3} Comparison of the wetting fractional flow
  ($F_w$) for 2D (left) and 3D (right) networks. Results from
  simulations with a maximum of 2 interfaces, case C (black symbols), 
  and simulations allowing up to 4 interfaces, case D (red symbols).  The
  straight line represents $F_w=S_w$, if both the fluids flow with
  same velocity.}
\end{figure}

Figure \ref{fig3} shows the wetting volumetric fractional flow rate
$F_w$ as a function of the wetting saturation $S_w$ for a maximum of 2 (C)
or 4 interfaces (D). The data for model D gives $F_w=S_w$ both in 2D and 3D.  
This indicates that when the maximum number of interfaces in the links is
higher, the fluids mix at the link level and they act as if there is
no viscosity contrast between them.  On the other hand, the maximum 
number of interfaces is small, the viscosity contrast it felt and the
least viscous fluids flow the fastest. 

%-----------------------------------------------------------------
%-----------------------------------------------------------------

Next, we turn to a lattice Boltzmann model which has no explicit
parameters for the bubble size or the number of interfaces and
permits arbitrary shapes of the fluid domains within the link. We 
base it on the original triangular lattice and the interaction rules 
first introduced by Gunstensen {\it et al}.\ \cite{grzz91}.  We implemented
the model on a $128\times 128$ biperiodic lattice. The size is rather
small, but suffices to simulate a representative $4 \times 4$
pipe-network, and is chosen in part to keep the Reynolds number
around 1 or smaller.  The capillary number Ca 
was larger than 9.  The pressure gradient was implemented as a constant 
body force in the diagonal direction.

%----------------------------------------------------------
\begin{figure}
\centerline{
  \includegraphics[width=0.35\textwidth,clip]{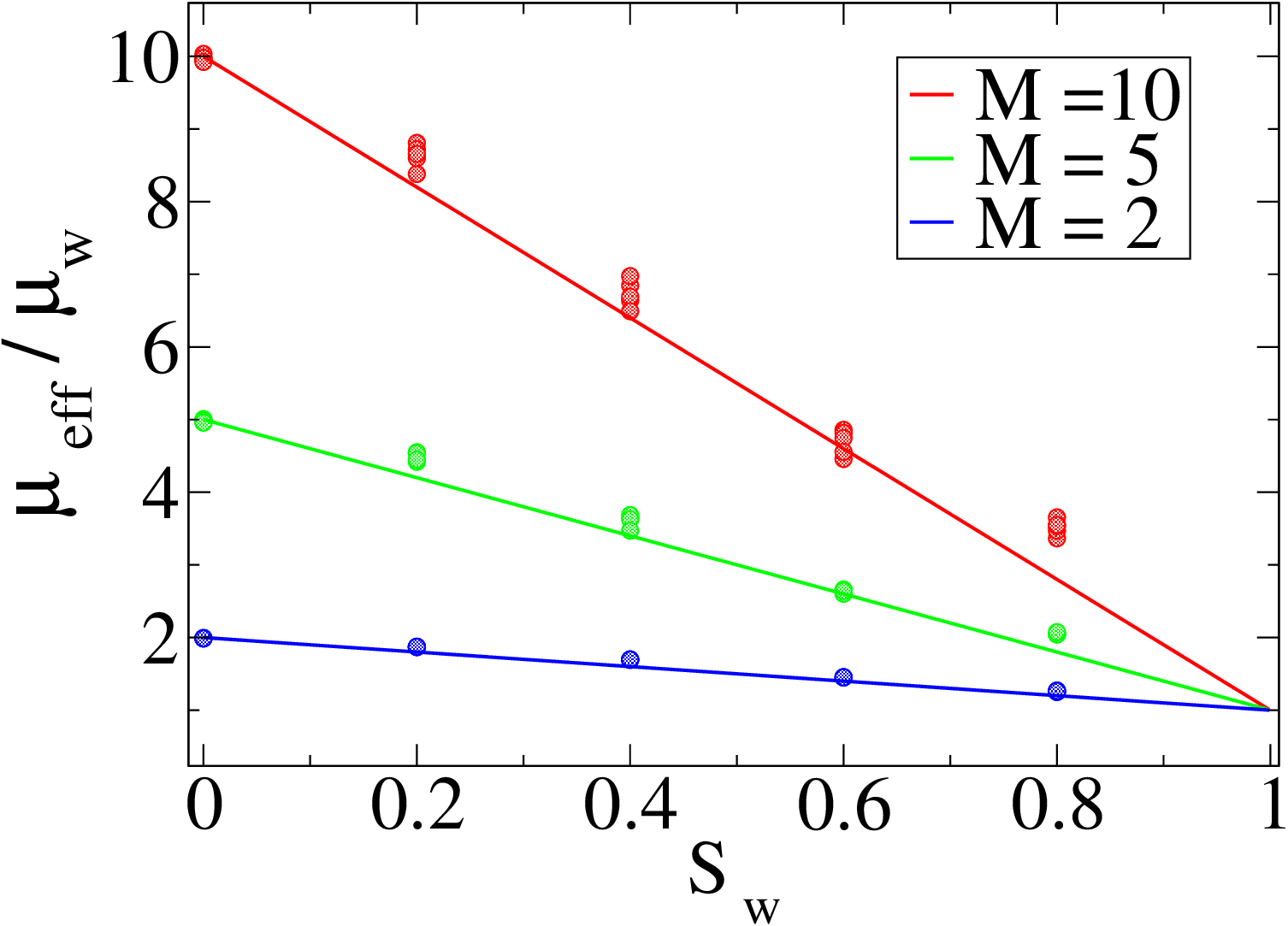}}
\caption{\label{fig4} The effective viscosity obtained
  from lattice Boltzmann simulations (dots) and compared with
  fits to equation (\ref{eq600}) with $\alpha=1$.}
\end{figure} 
%---------------------------------------------------------------

The simulations with a given
pressure gradient and measured flow rate were performed. In figure
\ref{fig4} the effective viscosity is plotted as a function of $S_w$.
The spread in the measurements at constant $S_w$ result from the
variability in the flow velocity corresponding to fluctuations in the
local viscosity values. The straight lines are
consistent with $\alpha = 1$ in equation (\ref{eq600}). This may
be the result of the large length to width ratio used for the links. 
The fluids
were found to organize in a way that combines both fluid elements in
parallel and in series, i.e.  there are both string- and plug-like
structures, see the supplementary material.

We have studied the effective viscosity of immiscible
two-fluid flow in porous media in the high capillary number limit
where the capillary forces may be ignored compared to the viscous
forces. We find that the Lichtenecker--Rother equation (\ref{eq600})
describes the effective viscosity well. The exponent 
depends on the fluid configuration, i.e.\ the number of
bubbles/interfaces in the pores. For small bubbles or many interfaces
in the pores, as with the Boltzmann model, we find $\alpha=1$, 
whereas when the bubbles are larger or the interfaces fewer in the 
pores, we find $\alpha=0.6$ in 2D (square and hexagonal lattices) and 
$\alpha=0.5$ in 3D for networks reconstructed from Berea sandstone 
and sand packs. The exponent $\alpha$ depends on the pore geometry.

\begin{acknowledgements}
The authors thank Dick Bedeaux, Carl Fredrik Berg, Signe Kjelstrup,
Knut J{\o}rgen M{\aa}l{\o}y, Per Arne Slotte and Ole Tors{\ae}ter for
interesting discussions. EGF and AH thank the Beijing Computational
Science Research Center CSRC and Hai-Qing Lin for hospitality. SS
was supported by the National Natural Science Foundation of China under
grant number 11750110430. This work was partly supported by the
Research Council of Norway through its Centers of Excellence funding
scheme, project number 262644.
\end{acknowledgements}

%--------------------------------------------------------------------

%--------------------------------------------------------------------
\end{document}